\documentstyle [prl,multicol,aps,psfig]{revtex}
\begin{document}
\title{Influence of measurement on the life-time and the line-width 
of unstable systems}
\author{Brahim Elattari$^{1,2}$ and S.A. Gurvitz $^3$\\
$^1$Departamento de f\'{\i}scica Te\'orica de la Materia Condensada, Universidad 
Aut\'onoma de Madrid, Cantoblanco, Madrid 28049, Spain\\
$^2$Universit\'e Choua\"\i b Doukkali, Fauclt\'e des Sciences,
El Jadida, Morocco\\
$^3$Weizmann Institute of Science, Department of Particle 
    Physics\\ 76100 Rehovot, Israel} 
\vspace{18pt}
\maketitle

\begin{abstract}
We investigate the quantum Zeno effect in the case of electron tunneling 
out of a quantum dot in the presence of continuous monitoring 
by a detector. It is shown that the Schr\"odinger equation for 
the whole system can be reduced to Bloch-type rate equations 
describing the combined time-development of the detector 
and the measured system. Using these equations we find that 
continuous measurement of the unstable system does not affect 
its exponential decay to a reservoir with a constant density 
of states. The width of the energy distribution of the tunneling electron, 
however, is not equal to the inverse life-time --
it increases due to the decoherence generated by the detector. 
We extend the analysis to the case of a reservoir 
described by an energy dependent density of states, and we 
show that continuous measurement of such quantum systems affects 
both the exponential decay rate and the energy distribution. 
The decay does not always slow down, but might be accelerated. 
The energy distribution of the tunneling electron may reveal the lines 
invisible before the measurement.  
\end{abstract}
\hspace{1.5 cm}  PACS: 73.23.Hk.03.65.Bz.73.23.-b
\vspace{18pt}
\begin{multicols}{1}
\section{Introduction}
 The interplay between quantum dynamics and quantum measurements 
has attracted much attention of physicists since the discovery 
of quantum mechanics. In fact, 
it was suggested that frequent or continuous observations of an 
unstable quantum system can inhibit or slow down its decay\cite{zeno}. 
This phenomenon is  known as 
the quantum Zeno effect. Usually this effect is associated 
to von Neumann's postulate in the theory of quantum 
measurements\cite{neu}.   
Indeed, in the classical example of two-level systems, the probability of 
a quantum transition from an initially occupied 
unstable state is given at short time, $\Delta t$, by  
$P(\Delta t)=a(\Delta t)^2$.
If we assume that $\Delta t$ is the measurement time, which 
consists in projecting the system onto the initial state, 
then after $N$ successive measurements the probability of 
finding the unstable system in its initial state, at time $t=N\Delta t$, is 
$Q(t)=[1-a(\Delta t)^2]^{(t/\Delta t)}$. 
It follows from this result that $Q(t)\to 1$ for $\Delta t\to 0$, 
i.e. suppression of quantum transition.  

During last years the Zeno effect has become a topic of
great interest. It has been discussed in the areas 
of radioactive decay\cite{panov1}, polarized light\cite{r1},
the physics of atoms\cite{r2,r3}, neutron physics\cite{r4}, 
quantum optics\cite{r5,r6} and mesoscopic physics\cite{gur1,hacken,eg}.
As a matter of fact, the theoretical and experimental efforts has been 
mainly concentrated on quantum transitions 
between isolated levels\cite{zeno1} characterized by an oscillatory 
behavior. In this latter case the slowing down 
of the transition rate has, indeed, been found. However, 
this was attributed to the decoherence generated by the detector 
without an explicit involvement of the projection 
postulate\cite{stod,gur1}. On the other hand, the slowing down of the 
exponential decay rate still remains a controversial 
issue, despite the fact that it is extensively 
studied\cite{Fonda,home,schul,kof,Panov} and further investigations
are clearly desirable.  
 
In this paper we focus our attention on observation of
spontaneous decay using a microscopic description  
of the measurement device (detector). The latter point 
should be very essential in any investigations of measurement problems. 
The quantum-mechanical description of the measurement device 
would allow us to study thoroughly the measurement process
without explicit use of the projection postulate, 
or introducing different phenomenological terms in quantum 
evolution of the measured system. Yet, the detector is a macroscopic 
system, the quantum mechanical analysis of which is very complicated. 
Thus one can expect that mesoscopic systems,
which are between the microscopic and macroscopic scales,
would be very useful for this type of investigation\cite{imry}. 
In addition, the actual experimental investigations of the quantum 
Zeno effect in mesoscopic systems are within reach 
of nowadays experimental techniques.

Our quantum-mechanical treatment of the entire system, 
including the detector is based on 
the new Bloch-type rate equations, which we derived here from  
the many-body Schr\"odinger equation. These equations would 
allow us to trace the quantum-mechanical behavior of 
the entire measured system during the measurement.
In contrast with the standard Bloch equations, our equations 
describe the reservoir states too. As a result, we can 
find the influence of the measurement on the energy distribution 
of the decayed system, as well as its time-evolution.  

The paper is organized as follows:
In Sect.~2 we gave a general quantum-mechanical description of an 
unstable system. We concentrated on conditions under which 
an exponential and non-exponential decay can be obtained, yet without 
including any measurement apparatus. The latter is introduced in 
Sect.~3 where
we derive the rate equations, describing the dynamical evolution of 
the measurement 
process. The results of our analysis are described in Sect.~4.
The last section is a summary. 
Some of the results described in this paper were presented in\cite{eg}.  

\section{Dynamics of a decayed system}
     
Let us consider a mesoscopic quantum dot coupled to an 
electron reservoir on its right, Fig.~1.
We assume only one level, $E_0$, inside the quantum dot, Fig.~1. 
The system is described by the following tunneling Hamiltonian 
written in the occupation number representation:
\begin{equation}
H_{QD} = E_0c_0^\dagger c_0
+\sum_\alpha E_\alpha c_\alpha^\dagger c_\alpha
+\sum_\alpha [\Omega_\alpha c_\alpha^\dagger c_0+H.c.] .
\label{a0}
\end{equation}
Here the operators $c_{\alpha}^+ (c_{\alpha})$ correspond to 
the creation 
(annihilation) of an electron in state $\alpha$ in the reservoir, 
and the operator $c_0^+ (c_0)$ is similarly defined for the state 
in the quantum dot. The $\Omega_\alpha$ are the hopping 
amplitudes between the states $E_0$, $E_\alpha$.
(We choose the gauge, where $\Omega_\alpha$ are real).   
This Hamiltonian (\ref{a0}) does not include any coupling 
with radiation or phonon fields as had been done for example   
in\cite{hepp,gur2}.
\begin{figure} 
\centering{\psfig{figure=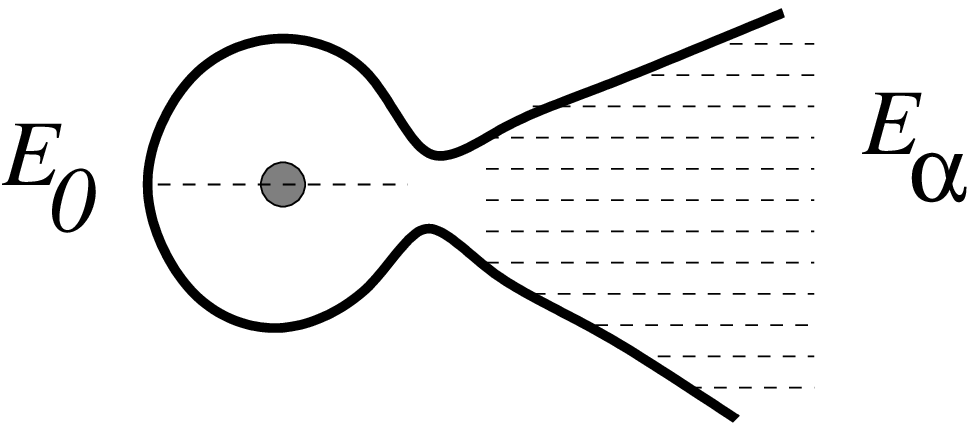,height=4cm,width=7cm,angle=0}}
{\bf Fig.~1:}
The energy level $E_0$ of the dot occupied by an electron,
which tunnels to continuous states $E_\alpha$ of the reservoir. 
\end{figure}

The initial state of the entire system $c^\dagger_0|0>$ 
corresponds to occupied quantum dot 
and empty states in the reservoir. 
This state is unstable: the Hamiltonian (\ref{a0}) requires it 
to decay to the continuum states $c^{\dagger}_{\alpha}|0>$ having 
the electron in the reservoir. The electron wave function, describing 
its evolution can be written in the most general way as  
\begin{eqnarray}
|\Psi (t)\rangle = \left [ b_0(t)c^\dagger_0 
+\sum_{\alpha} b_{\alpha}(t)c_{\alpha}^{\dagger} \right ]|0\rangle\ ,
\label{a1}
\end{eqnarray}
where $b_i(t)$ is the time-dependent probability amplitude to find 
the system in 
the corresponding state $i$ with the initial condition $b_0(0)=1$  
and $b_\alpha (0)=0$. Substitution of Eq. (\ref{a1}) into the 
time-dependent Schr\"{o}dinger equation 
$i\partial_t|\Psi(t)\rangle=H|\Psi(t)\rangle$, leads to an infinite set 
of coupled linear differential equations for the amplitudes $b(t)$. 
Applying the Laplace transform
\begin{eqnarray}
\tilde{b}(E)=\int_0^\infty b(t)\exp (iEt)dt
\label{aa1}
\end{eqnarray}
and taking into account the initial condition, we transform 
the differential equations for $b(t)$ into an infinite set 
of algebraic equations for the amplitudes $\tilde{b}(E)$:

\begin{mathletters}
\label{a2}
\begin{eqnarray}
&&(E-E_0)\tilde{b}_0(E)-\sum_{\alpha}\Omega_\alpha \tilde{b}_\alpha = i\ ,
\label{a2a}\\
&&(E-E_\alpha)\tilde b_\alpha (E)-\Omega_\alpha \tilde b_0(E) = 0\ .
\label{a2b} 
\end{eqnarray}
\end{mathletters}
In order to solve these equations we replace the amplitude 
$\tilde b_\alpha$ in Eq.(\ref{a2a}) by its expression obtained from 
Eq.(\ref{a2b}). Then we obtain
\begin{equation}
\left [ E-E_0-\sum_\alpha \frac{\Omega^2_\alpha }{E-E_\alpha} 
\right ]\tilde b_0(E)=i .
\label{a3}
\end{equation}   
Since the states in the reservoir are very dense, one can replace 
the sum over
$\alpha$ by an integral over $E_\alpha$. 
\begin{equation}
\sum_\alpha \frac{\Omega^2_\alpha }{E-E_\alpha} =
\int{\Omega^2 (E_\alpha )
\rho (E_\alpha )\over E-E_\alpha }dE_\alpha \, ,
\label{aa3}
\end{equation}   
where $\rho (E_\alpha )$ is the density of states in the reservoir. 
To evaluate this integral, we can split
the integral into its principal and singular parts,  
$-i\delta (E-E_\alpha)$. (Notice that $E\equiv E+i\epsilon$ in the 
Laplace transform, Eq.~(\ref{aa1})).   
As a result the original Schr\"odinger equation  
(\ref{a2}) is reduced to  
\begin{mathletters}
\label{a4}
\begin{eqnarray}
&&\left [E-E_0-\Delta (E)+i{\Gamma (E)\over 2}\right ]\tilde b_0(E)= i\, ,
\label{a4a}\\
&&\left (E-E_\alpha\right )\tilde b_\alpha (E)-\Omega (E_\alpha )
\tilde b_0(E) = 0 ,
\label{a4b}
\end{eqnarray}
\end{mathletters}   
where $\Gamma (E)=2\pi \rho(E)\Omega_\alpha^2(E)$ 
and $\Delta (E)$ is the  
energy-shift due to the principal part of the integral.

Using Eqs.~(\ref{a4}) and the inverse Laplace transform one 
can obtain the occupation probabilities of the levels 
$E_0$ and $E_\alpha$\cite{gur}. Yet, Eqs.~(\ref{a4}) are 
not convenient if we wish to include the effects  
of the environment. These effects can be determined in a natural way only in terms  
of the density matrix. For this reason  
we transform Eqs.~(\ref{a4}) to  
equations for the density matrix   
$\sigma_{ij}(t)\equiv b_i(t)b^*_j(t)$. The latter is directly related 
to the amplitudes $\tilde b(E)$ via the inverse Laplace transform
\begin{equation}
\sigma_{ij}(t) = \int_{-\infty}^{+\infty}\frac{dEdE'}
{(2\pi)^2}\tilde b_i(E)\tilde b^*_j(E')e^{i(E'-E)t} .
\label{a5}
\end{equation}
In order to proceed further with our derivations we have to  
know the $E$-dependence of $\Gamma$ and 
$\Delta$, determined by the density 
of states $\rho (E_\alpha )$ and the hopping amplitude 
$\Omega (E_\alpha )$, Eq.~(\ref{aa3}). We consider below
two important cases.

\subsection {Constant width}

Let us assume that $\Omega_\alpha^2 (E_\alpha )\rho (E_\alpha )$ 
is weakly dependent on the energy $E_\alpha$. As a result 
the width becomes a constant $\Gamma (E)=\Gamma_0$
and the energy shift $\Delta (E)$ tends to zero. 
In order to transform Eqs.~(\ref{a4}) to the equations 
for the density matrix by using the inverse Laplace transform (\ref{a5}),  
we multiply each of the Eqs.~(\ref{a4}) by the corresponding 
complex conjugate amplitude $b^*(E')$. 
For instance by multiplying Eq.~(\ref{a4a}) by $b_0^*(E')$ and 
subtracting the complex conjugated equation multiplied by $b_0(E)$
we obtain 
\begin{equation}  
(E'-E-i\Gamma_0)\tilde b_0(E)\tilde b_0^*(E') = 
-i[\tilde{b}_0(E)+ \tilde{b}_0^*(E')] ,
\label{a6}
\end{equation}   
It is quite easy to see that the inverse Laplace transform (\ref{a5})
turns this equation to the following one for the density matrix 
\begin{equation}
\dot \sigma_{00}(t) = -\Gamma_0\sigma_{00}(t)+[b_0(t)+b_0^*(t)]\delta(t) .
\label{a7}
\end{equation}
Notice that the initial time in the inverse Laplace transform 
(\ref{a5}) corresponds to $t=+0$, which allows us to ignore the term 
proportional to the $\delta$-function on the right hand side of Eq.~(\ref{a7}). 

Proceeding in the same way with all the other equations (\ref{a4}), we 
obtain the following set of equations for the density matrix 
$\sigma (t)$
\begin{mathletters}
\label{Reqn1}
\begin{eqnarray}
&&\dot \sigma_{00}(t) = -\Gamma_0\sigma_{00}(t) ,
\label{Reqn1a}\\
&&\dot \sigma_{\alpha\alpha}(t) = i\Omega_\alpha 
(\sigma_{\alpha 0}(t)-\sigma_{0\alpha}(t))
\label{Reqn1b}\\
&&\dot \sigma_{\alpha 0}(t) = i\epsilon_{0\alpha }
\sigma_{\alpha 0}(t)-i\Omega_\alpha \sigma_{00}(t)
-\frac{\Gamma_0}{2}\sigma_{\alpha 0}(t)\ ,
\label{Reqn1c}
\end{eqnarray}
\end{mathletters}  
with $\epsilon_{0\alpha }=E_0-E_\alpha$ and $\sigma_{0\alpha}
=\sigma^*_{\alpha 0}$.
 
It is interesting to compare Eqs.~(\ref{Reqn1}) 
with similar Bloch-type equations describing quantum transitions
between two isolated levels\cite{bloch,gur1}. 
In the case of the isolated levels ($E_1$ and $E_2$), the equations for
the density matrix $\sigma$ are symmetric with respect to $E_1$ and $E_2$. 
Whereas in the case of transition between the isolated 
($E_0$) and the continuum of states ($E_\alpha$)  
the corresponding symmetry, between $E_0$ and $E_\alpha$, is broken    
as can be seen, for example, in the equation for the off-diagonal 
term $\sigma_{\alpha 0}$ where the coupling 
with $\sigma_{\alpha\alpha}$ is missing. Eq.~(\ref{Reqn1c}).

Solving  Eqs.~(\ref{Reqn1}) we find 
the following expressions for the occupation probabilities, 
$\sigma_{00}$ and $\sigma_{\alpha\alpha}$, of the levels 
$E_0$ and $E_\alpha$, respectively\cite{gur}:
\begin{mathletters}
\label{Req1}
\begin{eqnarray}
&&\sigma_{00}(t) = e^{-\Gamma_0 t}\ ,
\label{Req1a}\\
&&\sigma_{\alpha \alpha}(t) =\frac{\Omega^2_\alpha}{\displaystyle 
(E_\alpha -E_0)^2+(\Gamma_0/2)^2}\nonumber\\
&&~~~~~~\times\left (1-2\cos (E_\alpha-E_0)t\,
e^{-\Gamma_0t/2}+e^{-\Gamma_0 t}\right )
\label{Req1b}  
\end{eqnarray}
\end{mathletters}
Notice that the line shape, 
$P(E_\alpha )\equiv \sigma_{\alpha \alpha}(t\to\infty )\rho$, 
given by Eq.~(\ref{Req1b}), 
is the standard Lorentzian distribution
\begin{equation}
P(E_\alpha ) =\frac{\Gamma_0/(2\pi )}{\displaystyle 
(E_\alpha -E_0)^2+(\Gamma_0/2)^2}
\label{loren}
\end{equation}
with the width $\Gamma_0$ corresponding to the inverse 
life-time of the quasi-stationary state, Eq.~(\ref{Req1a}). 

\subsection{Lorentzian density of states}

Let us consider the same problem studied before but with  
the decay width $\Gamma$ in Eq.~(\ref{a4a})
dependent on energy. This dependence could be generated either by 
the density of states in the reservoir 
$\rho (E_\alpha )$ or by  hopping amplitude, $\Omega_\alpha$,
 depending on $E_\alpha$. 
Here, we will restrict our selves to the frequent situation 
in which the density of states in the reservoir is modulated by 
a Lorentzian shape, as for instance by the resonant cavity, 
\begin{equation}
\rho(E_\alpha )=\bar\rho+ 
\frac{\Gamma_1/2\pi}{(E_\alpha-E_1)^2+\Gamma_1^2/4}.
\label{den} 
\end{equation}
Here $\bar\rho$ is a background component and $\Gamma_1$ is 
the width of the Lorentzian centered around $E_1$. 
Substituting Eq.~(\ref{den}) into Eq.~(\ref{aa3}) and assuming that 
the hopping amplitude $\Omega (E_\alpha)$ is weakly dependent on the 
energy $E_\alpha$, we can easily evaluate the integral (\ref{aa3}).
Then Eq.~(\ref{a4a}) becomes
\begin{equation}
\left (E-E_0+i\frac{\bar\Gamma}{2}-
\frac{\Omega_\alpha^2}{E-E_1+i\Gamma_1/2}\right )\tilde{b}_0(E)=i\ ,
\label{f}
\end{equation}
with $\bar\Gamma=2\pi\Omega_\alpha^2\bar\rho$\ .
Similar to the previous case we transform Eqs.~(\ref{a4}) into 
the rate equations for the corresponding density
matrix. For this reason we spilt Eq.~(\ref{f}) into 
two coupled equations by introducing an auxiliary amplitude
\begin{equation}
\tilde b_1 ={\Omega_\alpha\over E-E_1+i\Gamma_1/2} \tilde b_0. 
\label{fb}
\end{equation} 
As a result Eqs.~(\ref{a4}) become 
\begin{mathletters}
\label{gg}
\begin{eqnarray}
&&\left (E-E_0+i\frac{\bar\Gamma}{2}\right )\tilde{b}_0(E)
-\Omega_\alpha\tilde{b}_1(E) = i \\
\label{gg1}
&&\left (E-E_1+i{\Gamma_1\over 2}\right )\tilde{b}_1(E)
-\Omega_\alpha\tilde{b}_0(E)=0 \\
\label{gg3}
&&(E-E_\alpha)\tilde{b}_\alpha(E)-\Omega_\alpha\tilde{b}_0(E)=0
\label{gg2}
\end{eqnarray}  
\end{mathletters}
The amplitude $\tilde{b}_1$ can be interpreted as the probability 
amplitude of finding the electron in the state $E_1$
embedded in the reservoir. 
Now using the same procedure as that we applied for the derivation of 
Eqs.~(\ref{Reqn1}), we transform Eqs.~(\ref{gg}) into equations
for the density matrix $\sigma_{ij}(t)$:
\begin{mathletters}
\label{g1}
\begin{eqnarray}
\dot{\sigma}_{00}(t) &=& -\bar\Gamma\sigma_{00}+i\Omega_\alpha 
(\sigma_{01}-\sigma_{10})\\
\label{g1a}
\dot{\sigma}_{11}(t) &=& -\Gamma_1\sigma_{11}+i\Omega_\alpha 
(\sigma_{10}-\sigma_{01})\\
\label{g1b}
\dot{\sigma}_{01}(t) &=& i\epsilon_{10}\sigma_{01}
+i\Omega_\alpha (\sigma_{00}-\sigma_{11})
-\frac{\bar\Gamma+\Gamma_1}{2}\sigma_{01}\\
\label{g1c} 
\dot{\sigma}_{\alpha\alpha}(t) &=& i\Omega_\alpha
(\sigma_{\alpha 0}-\sigma_{0\alpha})\\
\label{g1d}
\dot{\sigma}_{0\alpha}(t) &=& i\epsilon_{\alpha 0}\sigma_{0\alpha}
+i\Omega_\alpha (\sigma_{00}-\sigma_{1\alpha})
-\frac{\bar\Gamma}{2}\sigma_{0\alpha}\\
\label{g1e}
\dot{\sigma}_{1\alpha}(t) &=& i\epsilon_{\alpha 1}\sigma_{1\alpha}
+i\Omega_\alpha (\sigma_{10}-\sigma_{0\alpha})
-\frac{\Gamma_1}{2}\sigma_{1\alpha} \, ,
\label{g1f}
\end{eqnarray}
\end{mathletters}
where $\epsilon_{ij} =E_i-E_j$ and $\sigma_{ji}=\sigma^*_{ij}$.  

Although Eqs.~(\ref{g1}) are quite different from  Eqs.(~\ref{Reqn1})  
obtained in the previous case, 
their interpretation is similar. In fact, Eqs.~(\ref{g1}) represent
Bloch-type equations describing decay of two-level system, as  
for instance an electron in the coupled double quantum dots, shown in Fig.~2. 
(The latter is described by Eqs.~(\ref{g1}) for $\bar\Gamma =0$). 
This is not surprising since the Lorentzian component of the density 
of states corresponds to a resonance state embedded in the reservoir.
\begin{figure} 
\centering{\psfig{figure=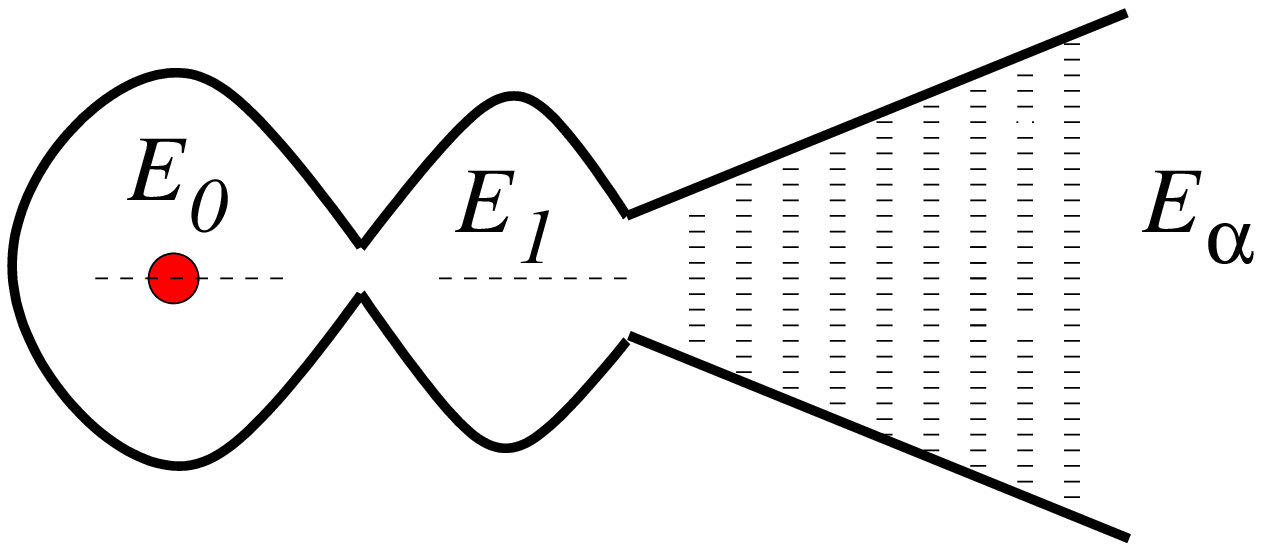,height=4cm,width=7cm,angle=0}}
{\bf Fig.~2:}
Double-dot system coupled to the reservoir of continuum states.
The electron is initially in the left dot.  
\end{figure}

Solving Eqs.~(\ref{g1}) (or Eqs.~(\ref{gg}))
one finds that 
\begin{equation}
\sigma_{\alpha\alpha}(t\to\infty)={\Omega_\alpha^2\left [(E_\alpha -E_1)^2
+\Gamma_1^2/4\right ]\over\displaystyle \left | \left (\epsilon_{\alpha 0}
+i{\bar\Gamma\over 2}\right)
\left(\epsilon_{\alpha 1}+i{\Gamma_1\over 2}\right )
-\Omega_\alpha^2\right |^2}
\label{bb1}  
\end{equation}
Thus, the line-shape $P(E_\alpha )\equiv\sigma_{\alpha\alpha}(t\to\infty)
\rho (E_\alpha)$ is not a pure  Lorentzian one, in contrast 
with the previous case, Eq.~(\ref{loren}). This is a reflection of 
non-exponential decay. The latter is generated by  
the energy dependence of the width $\Gamma (E)$ 
in Eq.~(\ref{a4a}), as for instance in the case of a  
 Lorentzian density of states. Indeed, by solving  
Eqs.~(18a)-(18c) for $\bar\Gamma =0$ we obtain
\begin{eqnarray}
&&\sigma_{00}(t)=\left |{\Gamma_1+\omega +2 i\epsilon_{01} \over 2\omega}
\exp\left (-{\Gamma_1-\omega\over 4}t\right )\right.\nonumber\\
&&\left.~~~~~~~~~~~~~~~ -{\Gamma_1-\omega +2 i\epsilon_{01} \over 2\omega}
\exp\left (-{\Gamma_1+\omega\over 4}t\right )\right |^2\ ,
\label{bb2}  
\end{eqnarray}
where $\omega =\sqrt{(\Gamma_1+2i\epsilon_{01})^2-16\Omega_\alpha^2}$. 
Hence, the decay is not a pure exponential one. In particular, 
it follows from Eq.~(\ref{bb2}) that the  
probability of finding the electron in the initial state for small $t$ is
$\sigma_{00}(t)= 1-\Omega_\alpha^2t^2$, 
in contrast with Eq.~(\ref{Req1a}), describing the exponential decay.  
Notice, that the absence of linear in $t$ term in 
$\sigma_{00}(t)$ is a prerequisition for the Zeno effect.

Consider Eq.~(\ref{bb2}) for $\Gamma_1\gg \Omega_\alpha, \epsilon_{01}$. 
One finds that the first exponential in Eq.~(\ref{bb2}) 
dominates the behavior of $\sigma_{00}(t)$ when $t$ increases. 
Thus 
\begin{equation}
\sigma_{00}(t)\simeq\exp \left (-{4\Omega^2_\alpha\over \Gamma_1}\ t\right )
~~~~{\mbox {for}}~~~ t\gg 2/\Gamma_1\, ,   
\label{bb3}  
\end{equation}
so that the decay becomes an exponential one. 
Respectively, the line-shape $P(E_\alpha )$, Eq.~(\ref{bb1}),    
becomes Lorentzian in the same limit:
\begin{equation}
P(E_\alpha )\simeq {1\over 2\pi}{4\Omega_\alpha^2/\Gamma_1
\over\displaystyle 
(E_\alpha-E_0)^2+\left ({2\Omega_\alpha^2\over\Gamma_1}\right )^2}
\label{OCC}
\end{equation} 

Using Eqs.~(18a)-(18c) one can also evaluate
the averaged decay-time of the electron in the state $E_0$, 
given by 
\begin{equation}
T=-\int_0^\infty t\ \dot{\sigma}_{00}(t)=\int_0^\infty 
\sigma_{00}(t)\ .
\label{time}
\end{equation}
The last integral can be evaluated directly from Eqs.~(18a)-(18c). 
As a result, we obtain for the decay-time 
\begin{equation}
T={\tau\over 1+\tau\bar\Gamma},
\label{ti}
\end{equation}
where 
\begin{equation}
\tau={1\over\Gamma_1}+
{4\epsilon_{10}^2+(\bar\Gamma+\Gamma_1)^2\over 
4\Omega_\alpha^2(\bar\Gamma+\Gamma_1)}\ . 
\label{ti1}
\end{equation}

In the case of zero background width, $\bar\Gamma =0$, the electron 
decay-time is 
\begin{equation}
T={1\over\Gamma_1}+
{4(E_1-E_0)^2+\Gamma_1^2\over 
4\Omega_\alpha^2\Gamma_1}\ . 
\label{ti2}
\end{equation}
Hence, instead of the intuitively expected answer $1/\Gamma_1$, 
we find an enhancement of the decay-time. This enhancement 
can be attributed to an oscillatory behavior of the 
electron between the states $E_0$ and $E_1$, which 
does not exist in the previous case of a constant 
width. In fact, it is even more surprising that 
for large $\Gamma_1$ the electron decay-time $T$ increases with $\Gamma_1$. 
Indeed, in the limit $\Gamma_1\gg \Omega_\alpha, \epsilon_{10}$
the decay-time becomes
$T\simeq \Gamma_1/4\Omega_\alpha^2$, in  accordance with 
Eqs.~(\ref{bb3}), (\ref{OCC}). Yet, in the opposite limit,  
$\epsilon_{10}\gg\Gamma_1$, the decay-time is $T\simeq 
(1+\epsilon_{10}^2/\Omega_\alpha^2)/\Gamma_1$. Thus it is inversely 
proportional to $\Gamma_1$, as expected. Such an unusual 
behavior of the decay-time can be understood in terms of a broadening 
of the level $E_1$ due to its coupling with continuum states $E_\alpha$,
Fig.~2. If $E_0=E_1$, this broadening results in an effective misalignment 
of the levels $E_0$ and $E_1$, thus destroying 
the resonant-tunneling condition. 
As a results the decay from the level $E_0$ to continuum slows down.
On the other hand, if $E_0\not =E_1$, the broadening of the level
$E_1$ would effectively diminishes the misalignment of the levels
$E_{0,1}$, so that the decay would be accelerated. 

It has been argued that the slowing down of the decay rate with 
$\Gamma_1$ in two stage processes, as shown in Fig.~2,
might be interpreted as the Zeno effect\cite{mensky}. This 
would imply that the electron decay from the level $E_1$ to the 
states $E_\alpha$ is treated as the measurement 
of the first transition from the level $E_0$ to the level $E_1$.
Such a ``measurement'' would localize the electron on the level $E_0$.
Yet, the electron in the reservoir is a part of the same measured 
system. Strictly speaking, it cannot be considered as an external 
system to itself. Actually the measurement must always imply  
an external macroscopic system interacting with the measured electron. 
Such a measurement device and its quantum mechanical description is
provided in the next section. 

\section{\bf Point contact as a detector of an unstable system}

As an example of the detector, monitoring decay of 
an unstable system, we consider a point contact placed near 
the quantum dot, Fig.~3. The point contact is connected with
two separate reservoirs. The reservoirs are filled up to the 
Fermi levels $\mu_L$ and $\mu_R$, respectively.  Therefore the 
current $I=e^2TV/2\pi$ flows from the left (emitter) to the right 
reservoir (collector),
where $T$ is the transmission coefficient of the point contact
and $eV=(\mu_L-\mu_R)$ is the bias voltage\cite{land}. 
(We consider the case of zero temperature). 
When the dot is occupied, Fig.~3a, the transmission coefficient of
the point contact decreases ($T'<T$) due to Coulomb repulsion generated by 
the electron in the dot. Respectively, the current through   
the quantum dot diminishes, $I'<I$. However, when the electron is in 
the reservoir, Fig.~3b, it is far away from the point contact. As a
result the transmission coefficient of the point contact becomes again
$T$, and the current increases. Thus, the point contact does  
represent a detector, which monitors the occupation of the quantum dot.
Actually, such a point contact detector has been successfully used 
in different experiments \cite{Buks}.  
Notice that the current variation ($I-I'$) can be a macroscopic
quantity if the applied voltage $V$ is large enough. 
\begin{figure} 
\centering{\psfig{figure=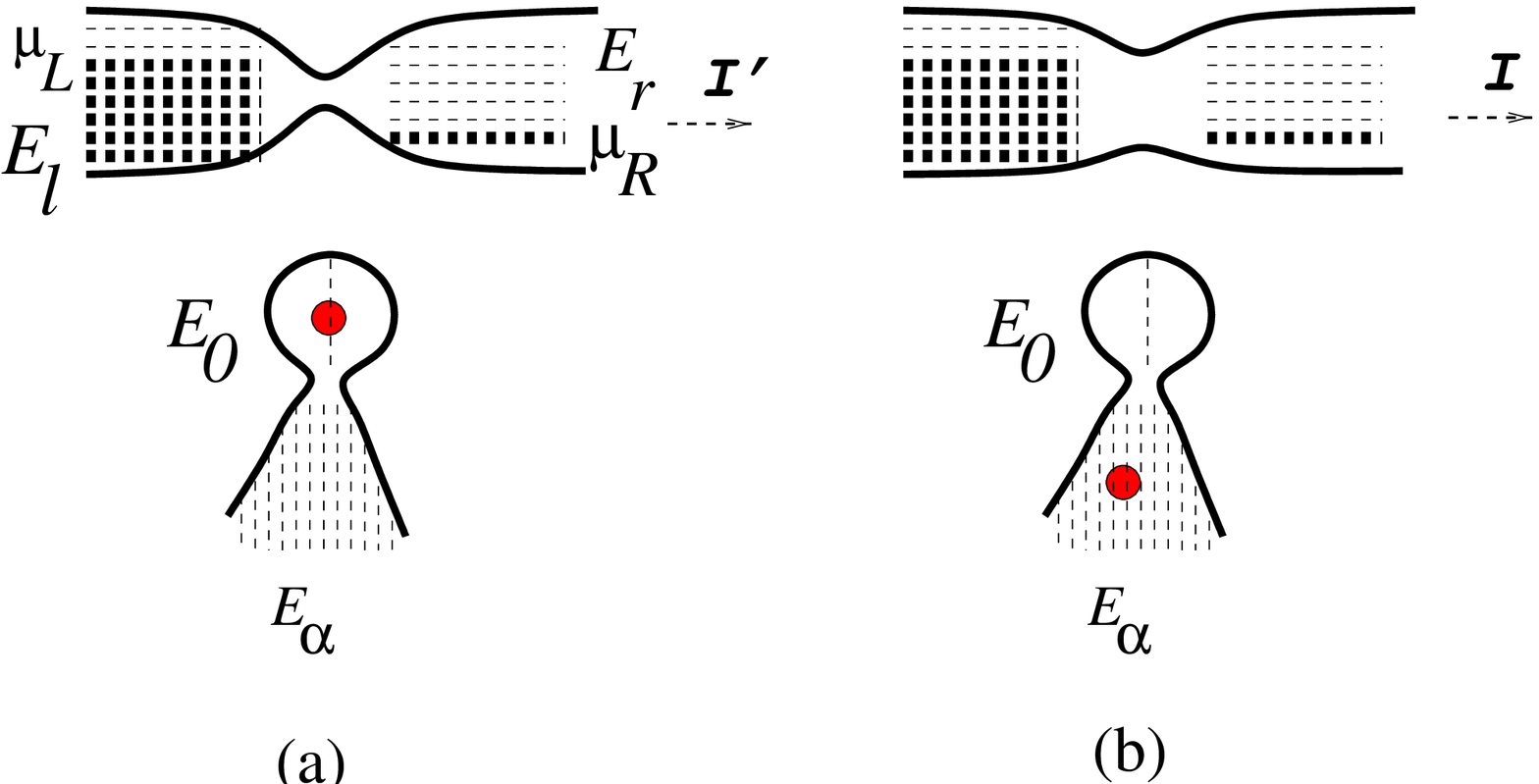,height=6cm,width=8.5cm,angle=0}}
{\bf Fig.~3:}
The point contact detector near the quantum dot. 
The energy level $E_0$ of the dot is occupied by an electron (a),
which tunnels from the state $E_0$, 
to continuous states $E_\alpha$ of the reservoir (b). $\mu_L$ and 
$\mu_R$ are the Fermi levels in the emitter and the collector, respectively. 
\end{figure} 

The dynamics of the entire system is determined by 
the many-body time-dependent Schr\"odinger equation 
$i\partial_t|\Psi(t)\rangle=H|\Psi(t)\rangle$, 
where the total Hamiltonian consists of three components 
$H=H_{QD}+H_{PC}+H_{int}$,
describing the quantum dot, the point contact detector, and 
their mutual interaction, respectively. The first component, $H_{QD}$,
is given by Eq.~(\ref{a0}). The two additional components,  $H_{PC}$ and 
$H_{int}$, can also be written in the form of 
a tunneling Hamiltonian as 
\begin{eqnarray}
H_{PC}&=&\sum_l E_l c_l^\dagger c_l+\sum_r E_r c_r^\dagger c_r
+\sum_{l,r} [\Omega_{lr} c_l^\dagger c_r+h.c.]
\nonumber\\
H_{int}&=&\sum_{l,r} [\delta\Omega_{lr} c_l^\dagger c_r+h.c.]
c_0^\dagger c_0\ ,
\label{bn1}
\end{eqnarray}
where the operators $c_{i}^+ (c_{i})$ correspond to the creation 
(annihilation) of an electron in state $i$. 
The $\Omega_{lr}$ are the hopping 
amplitudes between the states $E_l$, $E_r$ of the point contact. 
The quantity $\delta\Omega_{lr}=\Omega'_{lr}-\Omega_{lr}$ 
represents the variation of the point contact hopping amplitude, 
when the dot is occupied. 
    
Consider now the entire system in the initial condition corresponding 
to the occupied quantum dot and the reservoirs are   
filled up to the Fermi levels $\mu_L$ and $\mu_R$, Fig.~3a. 
We denote this state as $|\Psi (0)\rangle =c_0^\dagger|0\rangle$. This state 
is not stable: the Hamiltonian (\ref{a1}) requires it to decay to continuum
states having the form 
$c^\dagger_\alpha c^\dagger_ic^\dagger_j\cdots c_ic_j\cdots |0\rangle$.
In general, the total
wave-function at the time $t$ can be written as 
\begin{eqnarray}
&&|\Psi (t)\rangle = \left [ b_0(t)c_0^\dagger
+\sum_\alpha b_\alpha(t)c_\alpha^\dagger  
+\sum_{l,r} b_{lr}(t)c_r^{\dagger}c_l
\right.\nonumber\\
&&\left. ~~~~~~~~~~~~~~~~~~~~~~~~ 
+\sum_{\alpha ,l,r} b_{\alpha lr}(t)c_\alpha^{\dagger}c_r^{\dagger}c_l
+\cdots\right ]|0\rangle\ .
\label{b2}
\end{eqnarray}
Here $b(t)$ are the probability amplitudes of finding the system 
in the state
defined by the corresponding creation and annihilation operators.
As in the previous case, we substitute Eq.~(\ref{b2}) into the time dependent 
Schr\"odinger equation and use the
Laplace transform Eq.~(\ref{aa1}).
Then we find an infinite set of algebraic equations for the amplitudes 
$\tilde b(E)$. 
\begin{mathletters}
\label{b3}
\begin{eqnarray} 
&&(E-E_0)\tilde b_0-\sum_\alpha\Omega_\alpha\tilde b_\alpha 
-\sum_{l,r}\Omega'_{lr}\tilde b_{lr}=i\ ,
\label{b3a}\\
&&(E-E_\alpha)\tilde b_\alpha -\Omega_\alpha\tilde b_0
-\sum_{l,r}\Omega_{lr}\tilde b_{\alpha lr}=0\ ,
\label{b3b}\\
&&(E+E_l-E_r)\tilde b_{lr}-\sum_\alpha\Omega_\alpha\tilde b_{\alpha lr}
\nonumber\\
&&~~~~~~~~~~~~~~~~~~~
-\Omega'_{lr}\tilde b_0-\sum_{l',r'}\Omega'_{l'r'}\tilde b_{ll'rr'}=0
\label{b3c}\\
&&~~~~~~~~~~~~~~~~~~~~~~\cdots\cdots\cdots 
\nonumber
\end{eqnarray}
\end{mathletters}
These equations look much more complicated than the previous 
ones in Eqs.~(\ref{a4}),
describing the one-particle decay. Nevertheless, they can also be transformed 
into Bloch-type rate equations by tracing over the continuous degrees 
of freedom. 
Such a technique has been derived in
\cite{gur1,gur2,gur3}. In this paper we extend this technique 
by converting Eqs.~(\ref{b3}) into the rate equations, even without tracing  
over all continuum states\cite{eg}. As a results we obtain generalized 
Bloch-type equations that determine the energy distribution of the 
tunneling particles in the reservoirs, as well as the time-development
of the entire system. 

As in the previous case of the one-electron decay 
we replace the sums in Eqs.~(\ref{b3}) by the   
integrals, $\sum_k \to \int \rho (E_k)dE_k$.
Consider first the terms related to the left and the right 
reservoirs of the detector, Fig.~3.  
Let us assume that the corresponding hopping amplitudes, 
$\Omega_{lr}$, and the density of states $\rho(E_{l,r})$
are weakly dependent on the energies:
$\Omega_{lr}\equiv\Omega (E_l,E_r)=\Omega$, and $\rho(E_{l,r})=\rho_{L,R}$.
Let us also assume the large bias limit, $eV\gg \Omega^2\rho_{L,R}$,
so the integration limits can be extended to infinity. 
Then the integrations can be treated analytically.
Indeed, consider for instance Eq.~(\ref{b3a}). Replacing the amplitude
$\tilde b_{lr}$ by the corresponding expression obtained   
from Eq.~(\ref{b3c}), we find 
\begin{eqnarray}
&&\left (E-E_0-\int{{\Omega'}^2\rho_L\rho_RdE_ldE_r\over E+E_l-E_r}
\right )\tilde b_0(E)\nonumber\\
&&~~~~~~~~~~~~~~~~~~~~~~~~~~~~~~~~~~~-\sum_\alpha\Omega_\alpha\tilde b_\alpha
+{\cal F}=i\ ,  
\label{b33}
\end{eqnarray}
where ${\cal F}$ denotes the other terms where the amplitudes $\tilde b_0$ 
cannot be factorized out the integral. 
The integral in Eq.~(\ref{b33}) is treated in the same way as 
in Eq.~(\ref{aa3}) for one-electron tunneling. Namely, we split 
the integral into its principle value and singular part. 
The singular part yields $iD'/2$,
where  $D'=2\pi{\Omega'}^2\rho_L\rho_R eV=eT'V/(2\pi )$\cite{bardeen}, 
and the principal part is zero.

Consider now the non-factorizable terms ${\cal F}$ in Eq.~(\ref{b33}). 
These terms vanish in the large bias limit.
Indeed, all the singularities 
of the amplitude $\tilde{b} (E,E_l,E_{l'},E_r,E_{r'})$
in the $E_l, E_{l'}$-variables lie below the real axis. 
This can be seen directly from Eqs.~(\ref{b3}) 
by noting that $E$ lies above the real axis in the Laplace 
transform Eq.~(\ref{aa1}). Assuming that the transition amplitudes 
$\Omega$ as well as the density of states $\rho_{L,R}$ are independent of 
$E_{l,r}$, one can close the integration contour in the upper 
$E_{l,r}$-plane. Since the integrand decreases faster than $1/E_{l,r}$,
the resulting integral is zero. 

Applying analogous considerations to the other equations of the
system (\ref{b3}), we arrive at the following set of 
equations\cite{gur1,gur2} 
\begin{mathletters}
\label{b4}
\begin{eqnarray} 
&&\left (E-E_0+i{D'\over 2}\right )\tilde b_0
-\sum_\alpha\Omega_\alpha\tilde b_\alpha =i\ ,
\label{b4a}\\
&&\left (E-E_\alpha +i{D\over 2}\right )\tilde b_\alpha
-\Omega_\alpha\tilde b_0=0\ ,
\label{b4b}\\
&&\left (E+E_l-E_r+i{D'\over 2}\right )\tilde b_{lr} 
\nonumber\\
&&~~~~~~~~~~~~~~~~~~~~~~~~~~~~~~~
-\Omega' \tilde b_0-\sum_\alpha\Omega_\alpha\tilde b_{\alpha lr}=0\ ,
\label{b4c}\\
&&~~~~~~~~~~~~~~~~~~~\cdots\cdots\cdots 
\nonumber
\end{eqnarray}
\end{mathletters}
where $D=(2\pi )|\Omega_{lr}|^2\rho_L\rho_R(\mu_l-\mu_R)=eTV/(2\pi )$.

Proceeding in the same way with the remaining sum over the 
states $E_\alpha$ we finally arrive to the equations  

\begin{mathletters}
\label{bf4}
\begin{eqnarray} 
&&\left (E-E_0+i{D'\over 2}-{\cal I}(E)\right )
\tilde b_0=i\ ,
\label{bf4a}\\
&&\left (E-E_\alpha +i{D\over 2}\right )\tilde b_\alpha
-\Omega_\alpha\tilde b_0=0\ ,
\label{bf4b}\\
&&\left (E+\epsilon_{lr}+i{D'\over 2}-{\cal I}(E-\epsilon_{lr})\right )
\tilde b_{lr} 
-\Omega'\tilde b_0=0\ ,
\label{bf4c}\\
&&~~~~~~~~~~~~~~~~~~~\cdots\cdots\cdots 
\nonumber
\end{eqnarray}
\end{mathletters}
where $\epsilon_{lr}=E_l-E_r$ and 
\begin{equation}
{\cal I}(E)\equiv
\Delta (E)-i{\Gamma (E)\over 2}=\int{\Omega_\alpha^2 (E_\alpha )
\rho (E_\alpha )\over E-E_\alpha +iD/2}dE_\alpha \, ,
\label{bff4}
\end{equation}

Let us introduce the reduced density matrix of the entire system 
that includes the measured electron and the detector:
\begin{equation}
\sigma_{i,j}^{(n.n')}(t)=\langle\Psi (t)|n',j\rangle
\langle n,i| \Psi (t)\rangle\ , 
\label{bb4}
\end{equation}
where $|n,i\rangle$ denotes the state with $n$ electrons in the right 
reservoir, i.e. $\sum_r c^\dagger_r c_r |n\rangle=n|n\rangle$, and 
$i,j=\{0,\alpha \}$ denotes the state of the observed electron. 
This density matrix can be directly related to the amplitudes $b(t)$ of  
Eq.~(\ref{b2}):  
\begin{eqnarray}
&&\sigma^{(0,0)}_{00}(t)=|b_0(t)|^2,~~~
\sigma^{(0,1)}_{00}(t)=
\sum_{l,r}|b_0(t)b^*_{lr}(t)|^2,\nonumber\\
&&\sigma^{(1,1)}_{00}(t)=
\sum_{l,r}|b_{lr}(t)|^2,~~~
\sigma^{(0,1)}_{0\alpha}(t)=
\sum_{l,r,\alpha}|b_0(t)b^*_{\alpha lr}(t)|^2,\nonumber\\
&&~~~~~~~~~~~\ldots\ldots~~~~~~~~~~~~~,
~~~~~~~~~~~~~\ldots\ldots
\label{bb41}
\end{eqnarray}
 
Now we transform Eqs.~(\ref{bf4}) into equations for  
the density matrix $\sigma_{i,j}^{(n.n')}(t)$, without their explicit 
solution. It can be done by using the same procedure as in Sect. 2, 
for two cases, corresponding to weak and Lorentzian energy dependence of 
the function ${\cal I}(E)$, Eq.~(\ref{bff4}). Since we take for 
the definiteness the amplitude $\Omega_\alpha$ as independent 
of $E_\alpha$, these two cases correspond to constant or  
Lorentzian density of states $\rho (E_\alpha )$ in the reservoir. 
We demonstrate below that the influence of the measurement 
of the decay is distinctly different in both cases.  

\subsection {Constant density of states}

If the product $\Omega_\alpha^2\rho (E_\alpha )$ in  
Eq.~(\ref{bff4}) is weakly dependent on the energy $E_\alpha$, one 
can replace $\Gamma (E)=\Gamma_0$, and $\Delta (E)=0$ in 
Eqs.~(\ref{bf4})-(\ref{bff4}), where 
$\Gamma_0=2\pi\Omega_\alpha^2\rho$ = const.
In order to convert 
Eqs.~(\ref{bf4}) into equations for the density matrix, we  
we multiply each of these equations by the corresponding complex conjugate 
amplitude $b^*(E')$ and subtract the complex conjugated equation 
multiplied by $b_\alpha(E)$.
For instance Eq.~(\ref{bf4b}) becomes 
\begin{eqnarray}  
&&(E-E'+iD)b_\alpha(E)b_\alpha^*(E')\nonumber\\
&&~~~~~~~=\Omega_\alpha
[b_0(E)b_\alpha^*(E')-b^*_0(E')b_\alpha(E')]
\label{b5}
\end{eqnarray}
Then the inverse Laplace transform, Eq.~(\ref{a5}), converts 
Eq.~(\ref{b5}) to the following equation for the density matrix
\begin{equation} 
\dot{\sigma}_{\alpha\alpha}^{(0,0)}(t) =  
-D\sigma_{\alpha\alpha}^{(0,0)}(t)
+i\Omega_\alpha [\sigma_{\alpha 0}^{(0,0)}(t)
-\sigma_{0\alpha}^{(0,0)}(t)]\, .
\label{b6}
\end{equation}

Proceeding in the same way with all other equations (\ref{bf4}) and 
integrating over the continuum states of the collector and the emitter, we 
obtain the following infinite set of equations for the density matrix 
$\sigma (t)$
\begin{mathletters}
\label{Reqn}
\begin{eqnarray}
\label{Reqna}
\dot{\sigma}_{00}^{(n)} & = & -(\Gamma_0+D')\sigma_{00}^{(n)}
+D'\sigma_{00}^{(n-1)}\\
\label{Reqnb}
\dot{\sigma}_{\alpha\alpha}^{(n)} & = & -D\sigma_{\alpha\alpha}^{(n)}+
D\sigma_{\alpha \alpha}^{(n-1)}+i\Omega_\alpha(\sigma_{0\alpha}^{(n)}
-\sigma_{\alpha 0}^{(n)})\\
\label{Reqnc}
\dot{\sigma}_{\alpha 0}^{(n)} & = & i(E_0-E_\alpha)\sigma_{\alpha 0}^{(n)}
-i\Omega_\alpha \sigma_{00}^{(n)}
\nonumber\\
&&~~~~~-\frac{\Gamma_0+D+D'}{2}\sigma_{\alpha 0}^{(n)}
+\sqrt{DD'}\sigma_{\alpha 0}^{(n-1)}\, .
\end{eqnarray} 
\end{mathletters}
Here we denoted $\sigma_{\alpha 0}^{(n)}\equiv \sigma_{\alpha 0}^{(n,n)}$.
The initial conditions correspond to 
$\sigma_{ij}^{(n,n')}(0)=\delta_{i0}\delta_{j0}
\delta_{n0}\delta_{n'0}$.
Notice, that Eqs.~(\ref{Reqn}) involves only the diagonal 
density matrix elements    
with respect of the number of electrons in the collector 
$n$. This, however, does not imply the vanishing of the 
off-diagonal terms, $\sigma^{(n,n')}$. 
It means only their decoupling from the diagonal
terms in the equations of motion. This always takes place,
in transition between continuous states\cite{gur2,gur3}.  

Eqs.~(\ref{Reqn}) represent a generalization of the 
previously derived Bloch-type equations for quantum transport 
in mesoscopic systems\cite{gur1,gur2}. They have 
clear physical interpretation. 
Consider for instance Eq.~(\ref{Reqna}) for the probability 
of finding the electron inside the dot and $n$ electrons in the collector. 
This state decays due to one-electron hopping to the collector 
(with the rate $D'$), or due to the electron tunneling out of the dot 
(with the rate $\Gamma_0$). These processes are described by the first 
(``loss'') term in Eq.~(\ref{Reqna}). On the other hand, there exists 
the opposite (``gain'') process when the state with 
$(n-1)$ electrons in the collector converts into the state with $n$ 
electrons in the collector. It also takes place 
due to penetration of one electron through the point contact with the same 
rate $D'$. This process is described by the second term in Eq.~(\ref{Reqna}).  

The evolution of the off-diagonal density matrix elements 
$\sigma_{\alpha 0}^{(n)}(t)$ is given by Eq.~(\ref{Reqnc}). 
It can be interpreted in the same way as the rate equation for the 
diagonal terms. Notice, however, the difference between the ``loss'' 
and the ``gain'' terms. The latter can appear only 
due to coherent transition of the whole linear 
superposition\cite{gur1,gur2,gur3}. 

In order to determine the time-evolution of the observed electron 
we have to trace out over the detector states $n$ in Eqs. (\ref{Reqn}). 
As a result we obtain the following rate equations 
for the reduced density matrix $\sigma (t)\equiv \sum_n\sigma^{(n)}(t)$:
\begin{mathletters}
\label{eqn}
\begin{eqnarray}
\dot{\sigma}_{00} & = & -\Gamma_0\sigma_{00}
\label{eqna}\\
\dot{\sigma}_{\alpha\alpha} & = & i\Omega_\alpha(\sigma_{\alpha 0}
-\sigma_{0\alpha})
\label{eqnb}\\
\dot{\sigma}_{\alpha 0} & = & i(E_0-E_\alpha )\sigma_{\alpha 0}-
i\Omega_\alpha \sigma_{00}-\frac{\Gamma_0+\Gamma_d}{2}\sigma_{\alpha 0}\ .
\label{eqnc}
\end{eqnarray}
\end{mathletters}
Here $\Gamma_d=(\sqrt{D}-\sqrt{D'})^2$ is the 
decoherence rate, generated by the detector\cite{gur1}.

Let us compare these equations with Eqs.~(\ref{Reqn1}), describing  
decay of a single electron, which is not observed  
by the outside detector. We find the difference only in
equations for the  off-diagonal term,
Eqs.~(\ref{Reqn1c}) and (\ref{eqnc}), respectively.
The interaction with the detector
results in an increase of the damping term.
All other equations are unaffected by the measurement.
As a result, the probability of finding the system undecayed remains
the same as in the previous case, $\sigma_{00}(t)=\exp (-\Gamma_0t)$.
It means that the continuous monitoring 
of the unstable system does not slow down its decay rate. 
Nevertheless, it affects 
the energy distribution of the tunneling electron
$P(E_\alpha)\equiv \sigma_{\alpha\alpha}(t\to\infty )\rho$. 
Indeed, by solving Eqs.~(\ref{eqn}) in the limit of $t\to\infty$ we find 
a Lorentzian centered about $E_\alpha=E_0$:
\begin{equation}
P(E_\alpha)=
\frac{(\Gamma_0+\Gamma_d)/2\pi}{(E_0-E_\alpha)^2+(\Gamma_0+\Gamma_d)^2/4}\ .
\label{c1}
\end{equation}
If there is no coupling with the detector, $\Gamma_d=0$, 
the Lorentzian width (the line-width) $\Gamma_0$ is  
the inverse life-time of the quasi-stationary state, as 
given by Eq.~(\ref{loren}). 
However, as follows from Eq.~(\ref{c1}), the measurement results  
in a broadening of the line-width 
due to the decoherence generated by the detector. 
It now becomes $\Gamma_0+\Gamma_d$. 

On the first sight this result might look surprising. Indeed, it
is commonly accepted that the line-width does correspond 
to the life-time. Yet, we demonstrated here that it might not be the 
case, when the system interacts with an environment (the detector). 
To understand this result, one might think of the following argument. 
Due to the measurement, the energy level $E_0$ suffers an additional 
broadening of the order of  $\Gamma_d$. However, this broadening does 
not affect the decay rate of the electron $\Gamma_0$, 
since the exact value of $E_0$ relative to $E_\alpha$ is irrelevant 
to the decay process. In contrast, the probability distribution $P(E_\alpha)$ 
is affected because it does depend on the position of $E_0$ relative to 
$E_\alpha$ as it can be seen in
Eq.~(\ref{c1}).

Although our result has been proved for a specific detector, 
we expect it to be valid for the general case, provided that the density of 
states $\rho$ and the transition amplitude $\Omega_\alpha$ 
for the observed electron vary slowly with 
energy. This condition is sufficient to obtain a pure exponential decay 
of the state $E_0$.  On the contrary, if the product 
$\Omega_\alpha^2\rho (E_\alpha )$ depends sharply on energy $E_\alpha$, 
it yields strong $E$-dependence of $\Gamma$ and $\Delta$ in 
Eqs.~(\ref{a4}), (\ref{bf4}). 
This modifies both the exponential dependence of the decay
probability, $\sigma_{00}(t)$, and the energy distribution, $P(E_\alpha )$,
as that given by Eqs.~(\ref{bb1}) and (\ref{bb2}).
In particular, the decay probability for the small $t$ would be 
proportional to $t^2$. Therefore it is important to investigate the    
influence of the measurement in this case.

\subsection{Lorentzian density of states}

Consider for the definiteness that the density of states $\rho (E_\alpha )$
of the Lorentzian form, given by Eq.~(\ref{den}) and the 
amplitude $\Omega_\alpha$ is slowly dependent on the energy $E_\alpha$. 
In this case one obtains from Eq.~(\ref{bff4})
\begin{equation}
\Delta (E)-i{\Gamma (E)\over 2}=
-i{\bar \Gamma\over 2}+{\Omega_\alpha^2\over E-E_1+i(D+\Gamma_1)/2}\ .
\label{bc1}
\end{equation}
Substituting this result into
Eqs.~(\ref{bf4}) and introducing the auxiliary amplitudes
\begin{mathletters}
\label{cc1}
\begin{eqnarray}
\tilde b_1 & = & {\Omega_\alpha\over\displaystyle E-E_1
+i(D+\Gamma_1)/2} \tilde b_0
\label{cc1a}\\
\tilde b_{1lr}& = & {\Omega_\alpha\over E+E_l-E_r-E_1
+i(D+\Gamma_1)/ 2} \tilde b_{lr}
\label{cc1a}\\
& & ~~~~~\cdots\cdots
\end{eqnarray}
\end{mathletters}
we can rewrite Eqs.~(\ref{bf4}) in the following form 
\begin{mathletters}
\label{bc4}
\begin{eqnarray} 
&&\left (E-E_0+i{D'+\bar\Gamma \over 2}\right )\tilde b_0
-\Omega_\alpha \tilde b_1=i\ ,
\label{bc4a}\\
&&\left (E-E_1+i{D+\Gamma_1\over 2}\right )\tilde b_1
-\Omega_\alpha \tilde b_0=0\,
\label{bc4b}\\
&&\left (E-E_\alpha +i{D\over 2}\right )\tilde b_\alpha
-\Omega_\alpha\tilde b_0=0\ ,
\label{bc4c}\\
&&\left (E+E_l-E_r+i{D'+\bar\Gamma \over 2}\right )\tilde b_{lr} 
-\Omega' \tilde b_{1lr}=0\ ,
\label{bc4d}\\
&&~~~~~~~~~~~~~~~~~~~\cdots\cdots\cdots 
\nonumber
\end{eqnarray}
\end{mathletters}
where $\bar\Gamma =2\pi\Omega_\alpha^2\bar\rho$. The amplitudes 
$b_{1}(t),\ b_{1lr}(t),\ \ldots$ are the probability amplitudes 
of finding the electron on the resonant level $E_1$ of the continuum, 
for different numbers of electrons in the collector. 

It is quite clear that Eqs.~(\ref{bc4}) describe an unstable 
two-level system, which is continuously monitored by 
the point contact detector. If $\bar\Gamma =0$ the entire system 
is equivalent to that shown in Fig.~4, where the penetrability 
of the point contact is affected only when the observed electron 
occupies the level $E_0$.
 
\begin{figure} 
\centering{\psfig{figure=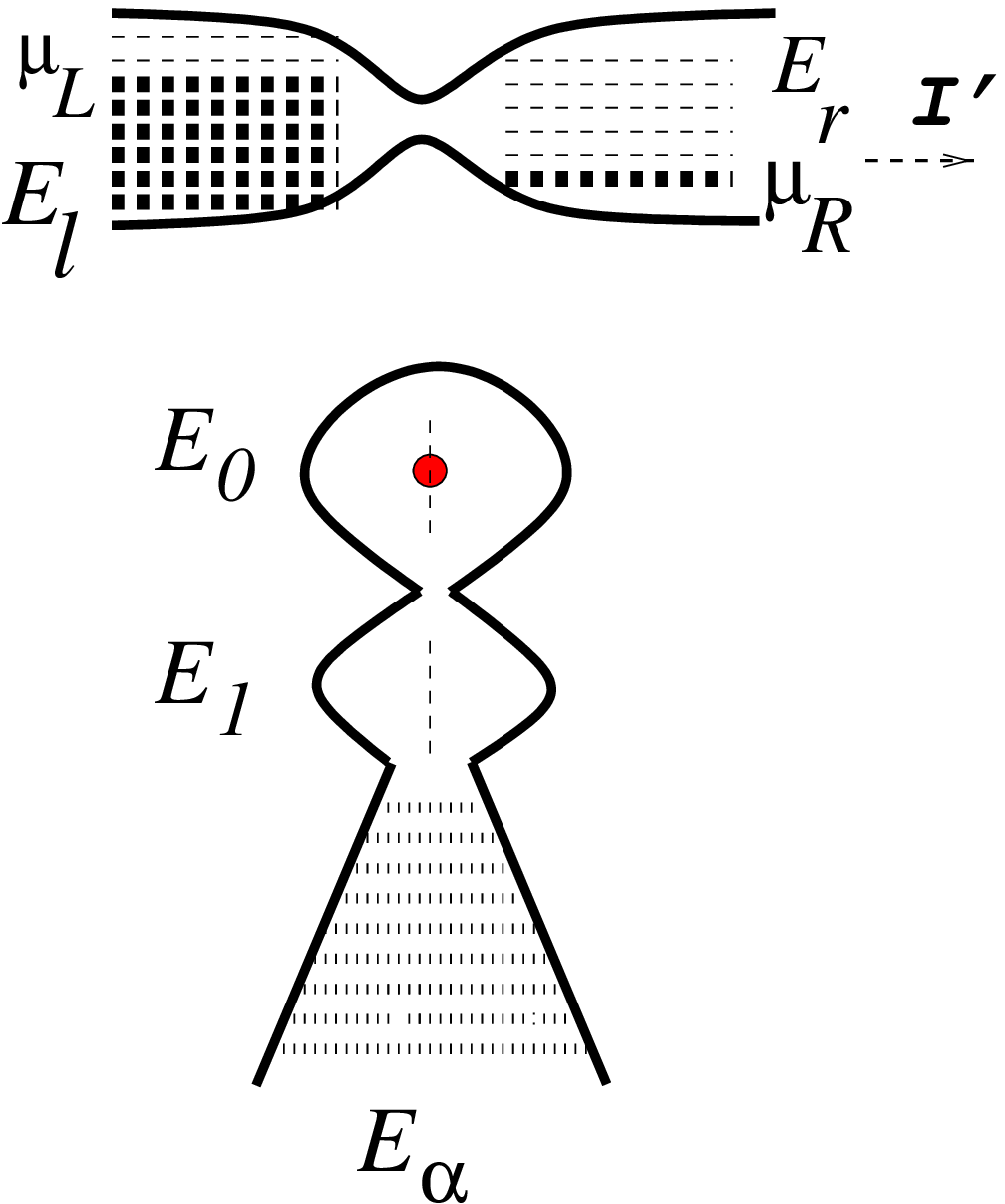,height=6cm,width=8.5cm,angle=0}}
{\bf Fig.~4:}
The point contact detector near the double-dot. 
The energy level $E_0$ of the first dot is initially 
occupied by an electron.
\end{figure} 

Using the same procedure as in the previous case we transform 
Eqs.~(\ref{bc4}) into equations for the density matrix $\sigma (t)$.
Finally we obtain the following set of equations, which 
involves only the diagonal terms in $n$ (similar to Eqs.~(\ref{Reqn})). 
For simplicity we consider only the case of $\bar\Gamma =0$.
\begin{mathletters}
\label{cn}
\begin{eqnarray}
\dot{\sigma}_{00}^{(n)} & = & -D'\sigma_{00}^{(n)}+D'\sigma_{00}^{(n-1)}
+i\Omega_\alpha (\sigma_{01}^{(n)}-\sigma_{10}^{(n)})
\label{cn1}\\
\dot{\sigma}_{11}^{(n)} & = & -(D+\Gamma_1)\sigma_{11}^{(n)}
+D\sigma_{11}^{(n-1)}\nonumber\\
&&~~~~~~~~~~~~~~~~~~~~~~~+i\Omega_\alpha (\sigma_{10}^{(n)}-\sigma_{01}^{(n)})
\label{cn2}\\
\dot{\sigma}_{01}^{(n)} & = & i\epsilon_{10}\sigma_{01}
+i\Omega_\alpha (\sigma_{00}^{(n)}-\sigma_{11}^{(n)})
\nonumber\\
&&~~~~~~~-\frac{D+D'+\Gamma_1}{2}\sigma_{01}^{(n)}
+\sqrt{DD'}\sigma_{01}^{(n-1)}
\label{cn3}\\
\dot{\sigma}_{\alpha\alpha}^{(n)} & = & -D\sigma_{\alpha\alpha}^{(n)}+
D\sigma_{\alpha \alpha}^{(n-1)}+i\Omega_\alpha(\sigma_{\alpha 1}^{(n)}
-\sigma_{1\alpha}^{(n)})
\label{cn4}\\
\dot{\sigma}_{0 \alpha}^{(n)} & = & i\epsilon_{\alpha 0}
\sigma_{0 \alpha }^{(n)}
+i\Omega_\alpha (\sigma_{00}^{(n)}-\sigma_{1\alpha }^{(n)})
\nonumber\\
&&~~~~~~~~-\frac{D+D'}{2}\sigma_{0\alpha }^{(n)}
+\sqrt{DD'}\sigma_{0\alpha }^{(n-1)}
\label{cn5}\\
\dot{\sigma}_{1\alpha}^{(n)} & = & i\epsilon_{\alpha 1}
\sigma_{1\alpha}^{(n)}+i\Omega_\alpha (\sigma_{10}^{(n)}
-\sigma_{0\alpha}^{(n)})
\nonumber\\
&&~~~~~~~~~~~~~~~~~~~~~-\frac{2D+\Gamma_1}{2}
\sigma_{1\alpha }^{(n)}+D\sigma_{1\alpha }^{(n-1)}.
\label{cn6}
\end{eqnarray} 
\end{mathletters}

These equations describe the microscopic behavior of 
the measured system and the detector at once. 
Their physical meaning is similar to that given for 
Eqs.~(\ref{Reqn}). In order to find the time-evolution of the 
observed electron alone we
trace out over the detector states $n$ in Eq.~(\ref{cn}). 
As a result, we obtain the following 
equations describing the time evolution of the corresponding 
reduced density matrix elements
\begin{mathletters}
\label{dd}
\begin{eqnarray}
\label{dda}
\dot{\sigma}_{00} & = & i\Omega_\alpha (\sigma_{01}-\sigma_{10})\\
\label{ddb}
\dot{\sigma}_{11} & = & -\Gamma_1\sigma_{11}
+i\Omega_\alpha (\sigma_{10}-\sigma_{01})\\
\label{ddc}
\dot{\sigma}_{01} & = & i\epsilon_{10}\sigma_{01}
+i\Omega_\alpha (\sigma_{00}-\sigma_{11})
-\frac{\Gamma_1+\Gamma_d}{2}\sigma_{01}\\
\label{ddd}
\dot{\sigma}_{\alpha\alpha} & = & i\Omega_\alpha(\sigma_{\alpha 0}
-\sigma_{0\alpha})\\
\label{dde}
\dot{\sigma}_{0\alpha} & = & i\epsilon_{\alpha 0}\sigma_{0\alpha}
+i\Omega_\alpha (\sigma_{00}-\sigma_{1\alpha})
-\frac{\Gamma_d}{2}\sigma_{0\alpha }\\
\label{ddf}
\dot{\sigma}_{1\alpha} & = & i\epsilon_{\alpha 1}\sigma_{1\alpha}
+i\Omega_\alpha (\sigma_{10}-\sigma_{0\alpha })
-\frac{\Gamma_1}{2}\sigma_{1\alpha }\ ,
\end{eqnarray} 
\end{mathletters}
where $\Gamma_d=(\sqrt{D}-\sqrt{D'})^2$ is the decoherence rate, generated 
by the detector. At first sight Eqs.~(\ref{dd}) resemble 
Eqs.~(\ref{g1}), where the background width $\bar\Gamma$ is replaced  
by the decoherence width $\Gamma_d$ generated by the detector.
Yet, $\Gamma_d$ does not enter the equations for diagonal 
density matrix elements, in contrast with the 
background width $\bar\Gamma$ that appears in Eq.~(\ref{g1a}).
This very essential difference follows from 
the fact that the measurement is a non-invasive one, and therefore
it does not result in additional dissipation processes.

Eqs.~(\ref{dd}) give a comprehensive description of the measured
system under continuous monitoring, including its energy  
distribution in the continuum, $P(E_\alpha)\equiv 
\sigma_{\alpha\alpha}(t\to\infty )\rho (E_\alpha)$. The latter can be obtained
from Eqs.~(\ref{dd}) without their explicit solution. Indeed, let us 
integrate each of Eqs.~(\ref{dd}) in the interval $0\leq t<\infty$ 
and take into account the initial condition 
$\sigma_{00}(0)=1$, $\sigma_{ij}(0)=0$ and $\sigma_{ij}(\infty)=0$,
except for $\sigma_{\alpha\alpha}(\infty)\not =0$. Then we obtain the 
following algebraic equations for $\bar\sigma =\int_0^\infty\sigma (t)dt$:
\begin{mathletters}
\label{dd1}
\begin{eqnarray}
\label{dd1a}
&&i\Omega_\alpha (\bar\sigma_{01}-\bar\sigma_{10})=-1\\
\label{dd1b}
&&-\Gamma_1\bar\sigma_{11}
+i\Omega_\alpha (\bar\sigma_{10}-\bar\sigma_{01})=0\\
\label{dd1c}
&&i\epsilon_{10}\bar\sigma_{01}
+i\Omega_\alpha (\bar\sigma_{00}-\bar\sigma_{11})
-\frac{\Gamma_1+\Gamma_d}{2}\bar\sigma_{01}=0\\
\label{dd1d}
&&i\epsilon_{\alpha 0}\bar\sigma_{0\alpha}
+i\Omega_\alpha (\bar\sigma_{00}-\bar\sigma_{1\alpha})
-\frac{\Gamma_d}{2}\bar\sigma_{0\alpha }=0\\
\label{dd1e}
&&i\epsilon_{\alpha 1}\bar\sigma_{1\alpha}
+i\Omega_\alpha (\bar\sigma_{10}-\bar\sigma_{0\alpha })
-\frac{\Gamma_1}{2}\bar\sigma_{1\alpha }=0\ ,
\end{eqnarray} 
\end{mathletters} 
where $\bar\sigma_{ji}=\bar\sigma^*_{ij}$. The 
energy distribution is $\sigma_{\alpha\alpha}(\infty )
\rho (E_\alpha )=2\Omega_\alpha {\mbox {Im}}\bar\sigma_{0\alpha}
\rho (E_\alpha )$.      

\section{Zeno and anti-Zeno effect}

In order to determine how the measurement affects the measured 
system, we solve Eqs.~(\ref{dd}), (\ref{dd1}) for $\Gamma_d\not =0$
and compare the solution with the corresponding one, obtained 
for $\Gamma_d =0$. Let us consider $\Gamma_1\gg\Omega_\alpha$, 
where the system displays an exponential decay, Eq.~(\ref{bb3}), 
for large enough $t$.

Let us first examine the probability 
of finding the system undecayed, $\sigma_{00}(t)$, 
given by Eqs.~(\ref{dda})-(\ref{ddc}). This quantity as 
a function of time (in the units $\Omega^{-1}_\alpha$) 
is shown in Fig.~5 in the logarithmic scale 
for $\Gamma_1/\Omega_\alpha=10$ and (a): 
$\epsilon_{10}=E_1-E_0=0$, (b):$\epsilon_{10}=10\Omega_\alpha$. 
\begin{figure} 
\centering{\psfig{figure=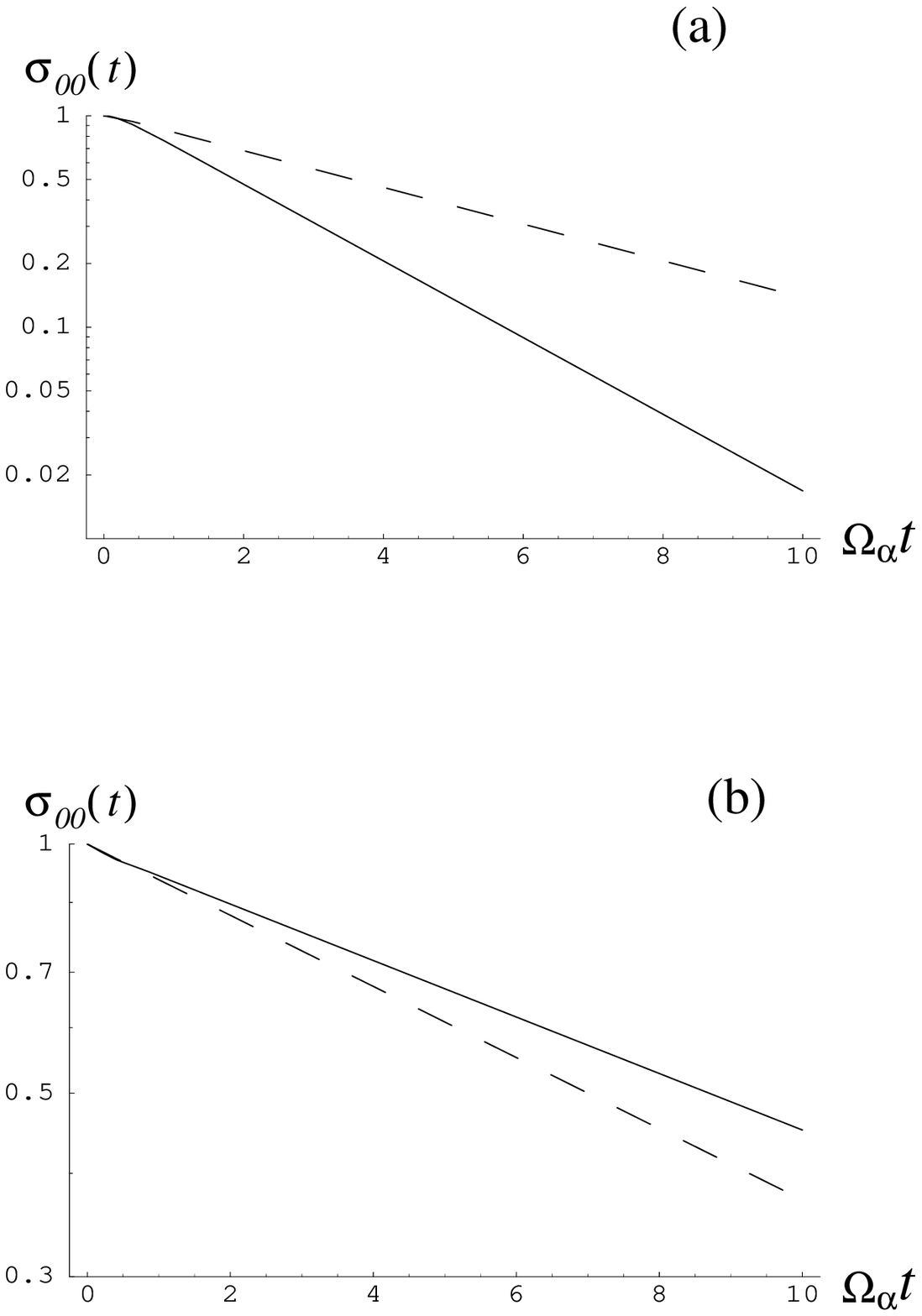,height=12cm,width=8.5cm,angle=0}}
{\bf Fig.~5:}
The probability of finding the electron inside the quantum dot
at the level $E_0$ (Fig.~4), where (a): $E_0=E_1$,
and (b): $E_1-E_0=10\Omega_\alpha$.
The solid line corresponds to $\Gamma_d=0$ (no measurement)
and the dashed line to  $\Gamma_d=10\Omega_\alpha$. 
\end{figure}
One finds from Fig.~5a that decay rate of the electron
monitored by the detector (the dashed line) slows down,
as expected from the Zeno effect. 
Yet, if the levels $E_0$ and $E_1$ are not aligned,  
the situation is different. It follows from Fig.~5b that 
the continuous monitoring of the decayed system leads 
to an acceleration of the decay, just the opposite to what is expected from Zeno effect 
(the anti-Zeno effect\cite{antizeno}). 
However, the later does not take place  at very 
short times, where the continuous observation still slows down 
the decay rate. This can be seen clearly in  Fig.~6, 
which magnifies the small $t$-region ($t<\Omega_\alpha^{-1}$) 
of Fig.~5b. 
\begin{figure} 
\centering
{\psfig{figure=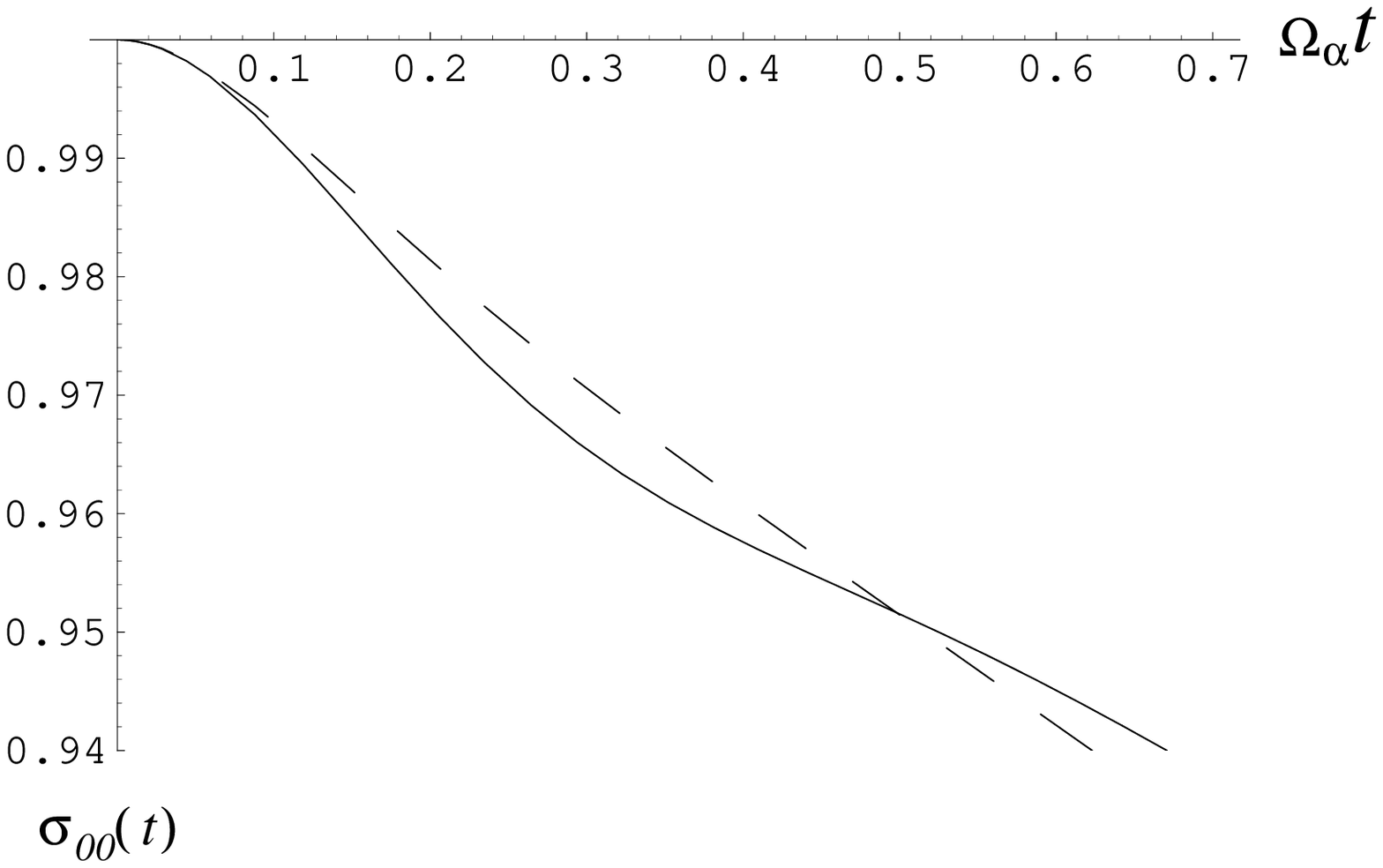,height=8cm,width=8.5cm,angle=0}}
{\bf Fig.~6:}
The probability of finding the electron undecayed 
for small $t$, and $(E_1-E_0)=10\Omega_\alpha$
All parameters are the same as in Fig.~5b. 
\end{figure} 

Actually, the Zeno and anti-Zeno effects  
can be revealed from the analytical 
solution of Eqs.~(\ref{dda})-(\ref{ddc}) for $t\gg\Omega_\alpha^{-1}$. 
Indeed, the behavior of $\sigma_{00}(t)$ in this region is dominated 
by the exponent with lesser exponential factor. The latter 
can be found directly from the secular determinant of 
Eqs.~(\ref{dda})-(\ref{ddc}) (and $\sigma_{10}=\sigma^*_{01}$)
in the limit of $\Gamma_1\gg\Omega_\alpha$. In which case one  obtains 
\begin{equation}
\sigma_{00}(t)\simeq \exp
\left (-{4(\Gamma_1+\Gamma_d)
\Omega_\alpha^2\over
4\epsilon_{01}^2+(\Gamma_1+\Gamma_d)^2}t\right )
\label{dd2}
\end{equation}
Therefore if $\epsilon_{01}\ll \Gamma_1+\Gamma_d$, the second 
term in the denominator of Eq.~(\ref{dd2}) dominates. As a result,
the decay rate slows down with the decoherence rate $\Gamma_d$
(Zeno effect). On the other hand, if $\epsilon_{01}\gg \Gamma_1+\Gamma_d$,
the first term dominates in the denominator of Eq.~(\ref{dd2})
and the decay accelerates with $\Gamma_d$ (anti-Zeno effect). 

One can arrive to the same conclusion by evaluating the average 
decay-time $T$ of the electron in the state $E_0$, Eq.~(\ref{time}). 
This can be obtained directly from Eqs.~(\ref{dda})-(\ref{ddc}).
In fact we find  
\begin{equation}
T=\frac{1}{\Gamma_1}+\frac{4\epsilon_{01}^2+(\Gamma_1+\Gamma_d)^2}
{4\Omega_\alpha^2(\Gamma_1+\Gamma_d)}.
\label{dd3}
\end{equation}
This is precisely the 
inverse exponential factor of Eq.~(\ref{dd2}) in the limit of 
$\Gamma_1\gg\Omega_\alpha$.

Consider now the energy distribution of the tunneling 
electron in the reservoir, $P(E_\alpha )=\sigma_{\alpha\alpha}(\infty)
\rho (E_\alpha )$, given by Eqs.~(\ref{dd1}). 
In the case of aligned levels, $\epsilon_{01}=0$, one finds:
\begin{equation}
P(E_\alpha )={2(\Gamma_1+\Gamma_d)(\Gamma_1\Gamma_d+4\Omega_\alpha^2)/\pi
\over 16\epsilon_{\alpha 0}^4+4\epsilon_{\alpha 0}^2(\Gamma_1^2+\Gamma_d^2
-8\Omega_\alpha^2)+(\Gamma_1\Gamma_d+4\Omega_\alpha^2)^2}
\label{dd4}
\end{equation}
If $\epsilon_{01}\not =0$, the analytical expression for $P(E_\alpha )$
is more complicated and less transparent. Therefore it is not presented here.

It follows from Eq.~(\ref{dd4}) that the width of the 
energy distribution 
does not correspond anymore to the inverse decay time, 
given by Eqs.~(\ref{dd2}), (\ref{dd3}). This stays in contrast 
with Eq.~(\ref{OCC}), where the line-width is precisely the 
inverse decay-time. Moreover the width of the energy
distribution, given by Eq.~(\ref{dd4}) always increases with 
the decoherence rate $\Gamma_d$, although the corresponding 
inverse decay-time decreases with $\Gamma_d$, 
Eqs.~(\ref{dd2}), (\ref{dd3}). Such a broadening of the energy 
distribution is shown explicitly in Fig.~7 
that displays $P(E_\alpha )$, Eq.~(\ref{dd4}), for $\Gamma_d=0$ 
(the solid line) and $\Gamma_d=10\Omega_\alpha$ (the dashed line).
Fig.~7 shows 
\begin{figure} 
\centering
{\psfig{figure=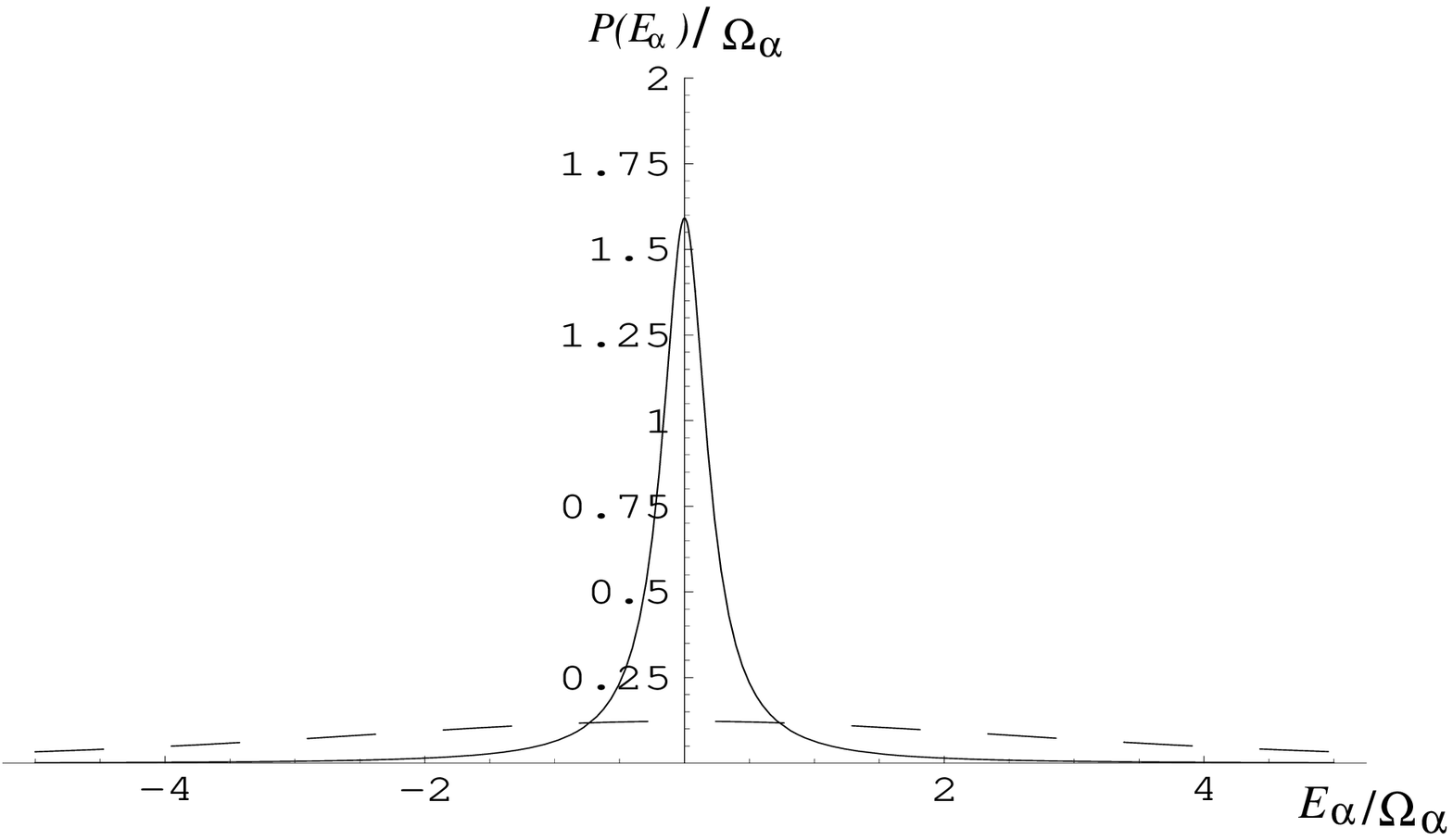,height=8cm,width=8.5cm,angle=0}}
{\bf Fig.~7:}
The energy distribution of the tunneling electron, $P(E_\alpha )$
for $E_1=E_0$, and $\Gamma_1=10\Omega_\alpha$.
The solid line correspond to $\Gamma_d=0$ and the dashed line to 
$\Gamma_d=\Gamma_1$. 
\end{figure}
\noindent 
that the width of the energy distribution increases very 
strongly with $\Gamma_d$, so that 
$P(E_\alpha )$ is almost flat for $\Gamma_d\gg\Omega_\alpha$.
In fact, the same strong broadening of the energy distribution  
takes place for $E_0\not =E_1$.

The described above measurement effects can be partially 
interpreted in terms of broadening of the level $E_0$ induced by  
the detector. One can expect that this broadening would always 
lead to spreading of the energy distribution. On the other hand,  
its influence on the decay rate depends 
whether the levels $E_0$ and $E_1$ are aligned or not. If $E_0=E_1$
the broadening of the level $E_0$ destroys
the resonant-tunneling condition, so that the decay to continuum 
slows down. However, if $E_0\not =E_1$, 
the same broadening would effectively diminish the levels displacement. 
As a result, the decay rate should increase. 
Yet, these arguments are not working at very short times,
when the decay rate slows down even for $E_0\not =E_1$,
as shown in Fig.~6. 

In general, such intuitive arguments, based only on the 
level broadening cannot explain all features of the 
energy distribution $P(E_\alpha )$, especially if the  
coupling of the level $E_1$ with the reservoir is weak. 
In this case the energy distribution, given by Eqs.~(\ref{dd1}) shows 
rather unusual dependence on the decoherence rate, $\Gamma_d$, generated 
by the detector. Let us take, for instance, $E_0=0$, $E_1=5\Omega_\alpha$
and $\Gamma_1=0.5\Omega_\alpha$.
The corresponding energy distribution $P(E_\alpha)$ is shown in Fig.~8
for $\Gamma_d=0$ (the solid line), $\Gamma_d=0.5\Omega_\alpha$ 
(the dashed line) and $\Gamma_d=10\Omega_\alpha$ (the dashed-dot line).
\begin{figure} 
\centering
{\psfig{figure=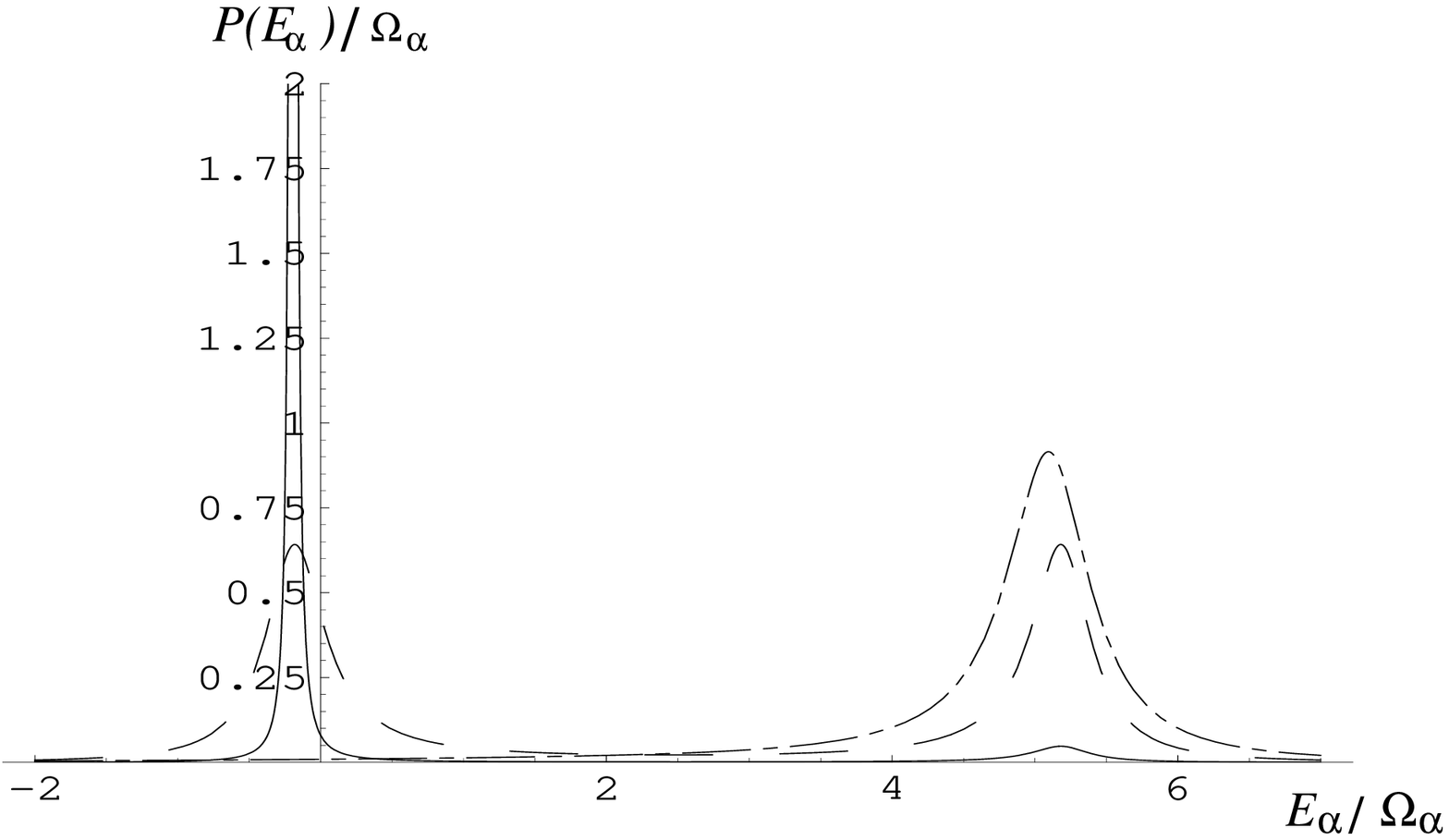,height=8cm,width=8.5cm,angle=0}}
{\bf Fig.~8:}
The energy distribution $P(E_\alpha )$ as given by Eq.~(\ref{dd1}) 
for $E_0=0$, $E_1=5\Omega_\alpha$, and $\Gamma_1=0.5\Omega_\alpha$.
The solid line correspond to $\Gamma_d=0$ the dashed line to 
$\Gamma_d=0.5\Omega_\alpha$, and the dashed-dot line to 
$\Gamma_d=10\Omega_\alpha$. 
\end{figure}
If there is no measurement ($\Gamma_d=0$) the energy distribution
$P(E_\alpha )$ is strongly peaked near $E_0$, whereas 
the second peak, at $E_\alpha\simeq E_1$, is almost invisible.  
Yet, by switching the detector on, the second peak increases with $\Gamma_d$.
In fact, both peaks are equally pronounced already for 
$\Gamma_d=\Gamma_1$. 
Then, for $\Gamma_d=20\Gamma_1$ the second peak becomes a dominant one 
(the dashed-dot line), whereas the first peak practically disappears.  

Thus we found that continuous monitoring of 
an electron in a double-dot system, weakly coupled 
with a reservoir changes completely the intensity 
of spectral lines. This can be partially interpreted in the 
following way. Consider first the case of no measurement,
$\Gamma_d=0$. Then a small coupling with the reservoir 
($\Gamma_1$) cannot essentially affect 
the eigenstates of the system. These are close to $E_{0,1}$,
providing that these levels are strongly detuned. Therefore 
the peak in the energy distribution of the tunneling 
electron lies near $E_0$, since the electron occupies this level. 
However, if we switch on the detector, it would decohere the 
electron inside the double-dot, before it tunnels to the 
reservoir. This means that electron tends to be equally 
distributed between the dots. Yet the second dot 
is directly coupled with the reservoir, whereas the first dot is not, 
(Fig.~4). As a result, the energy distribution would display the 
peak near $E_1$.

\section{summary}

We presented detailed quantum-mechanical analyses 
of a decayed quantum system under continuous monitoring. As an 
example we considered an electron tunneling 
from a discrete level of quantum dot to the empty reservoir. The essential 
point in our analysis is a full quantum mechanical account of the 
macroscopic (mesoscopic) detector. This allowed us to investigate 
quantum Zeno effect as generated by the Schr\"odinger evolution 
of the entire system, without invoking the projection postulate.  
In contrast with the previous treatments, which concentrate only
on the time-evolution, we analyzed here the complementary energy distribution 
of the decayed system, as well.   

The results of our analysis clearly demonstrate that the evolution 
of a decayed system under continuous monitoring is crucially 
related to the energy dependence of the density of states 
in the reservoir and 
the tunneling amplitude. If these quantities are weakly 
dependent on the energy, a pure exponential decay of 
the quantum system takes place. 
In this case the measurement does not affect the decay rate. Yet,
the energy distribution of the electron in the reservoir 
becomes strongly broadened. As a result, the corresponding line-width 
is no more given by the inverse decay time. 

In the case of strong energy dependence of the
density of states (or of the tunneling amplitude) 
the situation is different. As an example we considered 
the Lorentzian component in the density of states of   
the width greater than the tunneling amplitude.  
In this case the decay is not an exponential one for small times. But it  
becomes exponential when the time 
increases. Nevertheless, in contrast with the previous case 
the measurement affects the decay rate,
even in the exponential regime. The effect, however, 
depends on displacement between the initial state energy 
and the mid-energy of the Lorentzian density of states. 
If these energies are close to each other,  we obtain the expected 
Zeno effect, i.e. the slowing down of the decay rate at any time. 
Yet, in the opposite case, when the displacement of these 
energies is larger than the Lorentzian width, the measurement 
induces an acceleration of the decay rate  
in the regime of the exponential decay (the anti-Zeno effect).
At small times, however, the measurement always slows
down the decay rate.

With respect to the energy distribution of the tunneling particle, 
the measurement always generates strong broadening. 
However, for very narrow Lorentzian shape of the density of states
(the Lorentzian width is smaller than the tunneling amplitude)
the effect of the measurement is more peculiar. 
This takes place where the displacement between the initial energy 
and the Lorentzian mid-energy is large. Then, if there is no measurement, 
the energy distribution peaks near the initial state energy, whereas 
the second peak (near the Lorentzian center) is very weak. 
However, when the measurement 
is switched on, the peak at the initial energy disappears with 
the increase of decoherence, generated by the detector.
On the other hand, the second peak near the Lorentzian mid-energy 
arises. 

All these results were obtained without taking into account the coupling of 
the fermionic 
quantum dots to the radiation or to the phonon fields. In fact, in a weak coupling
regime our treatment can be extended to this case. Yet, we do not  
expect any essential modification of the results obtained in this paper.   
The strong coupling regime, however, needs a special consideration. 

Our analysis shows different effects generated by continuous 
measurement. All of them were obtained solely from the time-dependent 
Schr\"odinger equation applied to the entire system. 
Moreover, it looks that some of these effects might be in a contradiction 
with the projection postulate, which leads to the slowing 
of the decay rate. Yet, this this point deserves special 
investigation on the level of the macroscopic description of the 
measuring device. It would involve a proper definition 
of the measurement time, needed for application of the projection 
postulate. Although we are not coming to this point in our paper,
we nevertheless consider our analysis as a necessary step 
for a better understanding of the measurement problem and 
the nature of the projection postulate.

\section{Acknowledgments}
We thank  Y. Imry, S. Levit, A. Kofman and G. Kurizki,
for valuable discussions.
One of us (S.G.) would like to acknowledge the hospitality of 
Oak Ridge National Laboratory and 
TRIUMF, while parts of this work were being performed. One of us 
(B.E.) gratefully acknowledge the support of GIF and the
Israeli Ministry of Science and Technology and the French Ministry of
Research and Technology

\end{multicols}
\end{document}